# PRIME-DP: Pre-trained Integrated Model for Earthquake Data Processing.

Ziye Yu[1], Yuqi Cai[1], Weitao Wang[1], Yanru An[2], Lu Li[1], Yueyang Xia[1], Yunpeng Zhang[1]

1. Institute of Geophysics, China Earthquake Adminstration, Beijing 100081, China.

2. China Earthquake Networks Center, Beijing 100081, China.

yuziye@cea-igp.ac.cn

Abstract: We propose a novel seismic wave representation model, namely PRIME-DP (Pre-trained Integrated Model for Earthquake Data Processing), specifically designed for processing seismic waveforms. Most existing models are designed to solve a singular problem. Unlike these models, PRIME-DP is capable of multi-task single station seismic waveform processing, including Pg/Sg/Pn/Sn phase picking and P polarization classification. Moreover, it can be fine-tunned to various tasks, such as event classification without architecture modifications. PRIME-DP can achieve a recall rate of over 85% for Pg and Sg phases on continuous waveforms and achieves over 80% accuracy in P polarization classification. By fine-tuning classification decoder with NeiMeng dataset, PRIME-DP achieves 95.1% accuracy on event.

## 1. Introduction

The incorporation of deep neural network (DNN) technology into the geophysical approaches has demonstrated advantages in precision, accuracy, and efficiency in computational-expensive calculations, compared to conventional workflows with manual interference and human bias (Beroza et al., 2021; Mousavi and Beroza, 2022). Typical applications include seismic phase picking (Zhu and Beroza, 2019; Ross et al., 2018a; Wang et al., 2019; Mousavi et al., 2020; Xiao et al., 2020; Yu and Wang, 2022), phase association (Ross et al., 2019; McBrearty et al., 2019a,b; Yu and Wang, 2022), earthquake detection (Yang et al., 2021; Yano et al., 2021), P phase polarity determination (Ross et al., 2018b; Hara et al., 2019; Tian et al., 2020; Zhao et al., 2023), focal mechanism solution inversion (Kuang et al., 2021; Li et al., 2023), seismic waveform denoising (Zhu et al., 2019; Wang et al., 2021; Yang et al., 2022; Novoselov et al., 2022), earthquake localization (DeVries et al., 2018; Lomax et al., 2019; Mousavi and Beroza, 2019; Zhang et al., 2020; Zhang et al., 2022; McBrearty and Beroza, 2023), and earthquake classification (Linville et al., 2019; Bregman et al., 2021; Kong et al., 2020). Currently, most neural network models in seismology focus on a single task, such as phase picking, which takes seismic waveforms as input information and output P- and S-waves arrival times of a given seismic event (Zhu et al., 2019; Ross et al., 2018a; Wang et al., 2019; Mousavi et al., 2020; Xiao et al., 2020; Yu and Wang, 2022). Some studies have attempted to incorporate more information into the input, for instance, Munchmeyer et al. (2021) explored earthquake localization and magnitude estimation simultaneously, and SeisCLIP (Si et al., 2024) combined phase arrival times, epicentral distances, and azimuths into a single neural network to create a model capable of processing a broader range of information.

Seismological data analysis involves processing large volumes of data, with different types of data for each analysis task. Training DNN models for each task will require vast amounts of manually annotated data. However, for specific task like tele seismic phases detection, the availability of manually annotated data is relatively limited. Consequently, transfer learning has been applied in seismology. Transfer learning refers to the models trained on extensive datasets applied to new tasks with few labeled samples training. This approach has demonstrated effective performance in image and text processing and is similarly utilized in seismology to reduce the reliance on extensive training datasets. Studies indicate that models trained with large datasets perform well in new scenarios with limited data, such as seismic event classification.

Models trained on substantial data for transfer learning tasks are known as pre-trained models. The unidirectional models, represented by the GPT series (Radford et al., 2018; Radford et al., 2019; Brown et al., 2020), consider only the preceding information in sequence and are typically used in text generation, and multimodal tasks involving text, images, and audio. In contrast, bidirectional models, exemplified by the BERT (Devlin et al., 2018), consider the full sequence in text processing. Generative models, which can now handle up to hundreds of billions of trainable parameters, have shown excellent performance in text generation. SeisCLIP utilizes generative models to process seismic spectra from multiple stations and generate required information. However, seismological research often necessitates the analysis of raw waveforms to extract more accurate information, where bidirectional models typically achieve higher accuracy (Yu et al., 2023).

In 2023, the China Seismological Network Center released CSNCD dataset, which contains over 45 million manually annotated information including over 80 type of phases, P polarity, earthquake types, and magnitudes. Counting on this, we developed a bidirectional neural network pre-trained model for single-station data analysis, named the China Seismic Bidirectional Multi-Decoder Model. Compared to unidirectional generative models, our bidirectional model offers higher output accuracy and is better suited for waveform analysis. Our model takes raw waveform data as input and employs multi-layer convolution neural network (CNN) to extract features from the raw waveforms. To further process these features, we incorporated a bidirectional Transformer architecture that utilizes multiple decoders for many tasks such as phase detection, initial motion detection, and earthquake type classification. As a pre-trained model, PRIME-DP includes outputs for four phase types (Pg, Sg, Pn, Sn), P-wave initial motion direction, earthquake types, and raw waveforms, with the raw waveform output being self-supervised. After training, modifications to the encoder or feature vectors can adapt the model for other tasks, indicating that the feature vectors generated by PRIME-DP contain comprehensive information suitable for seismic analysis. Tests shows that our model not only accomplishes conventional tasks such as Pg, Sg, Pn phase detection, P-wave initial motion direction determination, and event type classification but also adapts through transfer learning to seismic event classification.

# 2. Method

## 2.1 Training dataset

We utilize the CSNCD dataset from the China Seismological Network Center to train and evaluate our model. The CSNCD dataset is a large-scale global seismic signal labeled dataset, distinguished by its coverage of diverse geographical areas and various geological environments. It includes over 1.3 million seismic events distributed worldwide. The CSN stations are located in China mainland (Figure 1a blue triangle) and the events (Figure 1a black dot) are within 2000km from the stations.

We have divided the CSNCD dataset into training, validation, and test datasets, assigning data from 2009 to 2019 to the training set, data from 2020 to the validation set, and data from 2021 and 2022 to the test set. The training and validation sets are used for training, fine-tuning parameters, and model selection, while the test set is solely for evaluating the final performance and results of the model. In terms of enhancing model robustness, the CSNCD dataset includes a wide range of signal-to-noise ratios (SNR) and epicenter distance (lower than 2000km) (Figure 3 and Figure 4), which enables our model to effectively learn and predict on data of varying quality, thereby improving its adaptability and generalization capability.

We selected four phases (Pg, Sg, Pn and Sn) (Figure 1c), P polarity (Figure 1d, C/U for up and R/D for down), event type (Figure 2) as targets for our model. As shown in Figure 5, the input of our model $x \in \mathbb{R}^{3\times N}, N = 10240$ is seismic waveforms for the vertical (Z), north (N), and east (E) components, which is 10240 sampling points at 100Hz. We minimize the processing of waveforms and only include normalization.

Phase labels $d^{phase} \in \mathbb{R}^{5\times N}$ reflect the phase probability in the waveform data, with the black line representing the ambient noise level, and blue, red, yellow, and green lines respectively representing the probability distributions of Pg, Sg, Pn, and Sn phases. P polarity in the CSNCD dataset (Figure 1d) are labeled as long as waveform $d^{polar} \in \mathbb{N}^N$. We selected the 10.24 seconds time window centered with Pg to output the classification of P polarity. The label of type of events is $d^{type} \in \mathbb{N}^7$. Most of the events are earthquake (Figure 2).

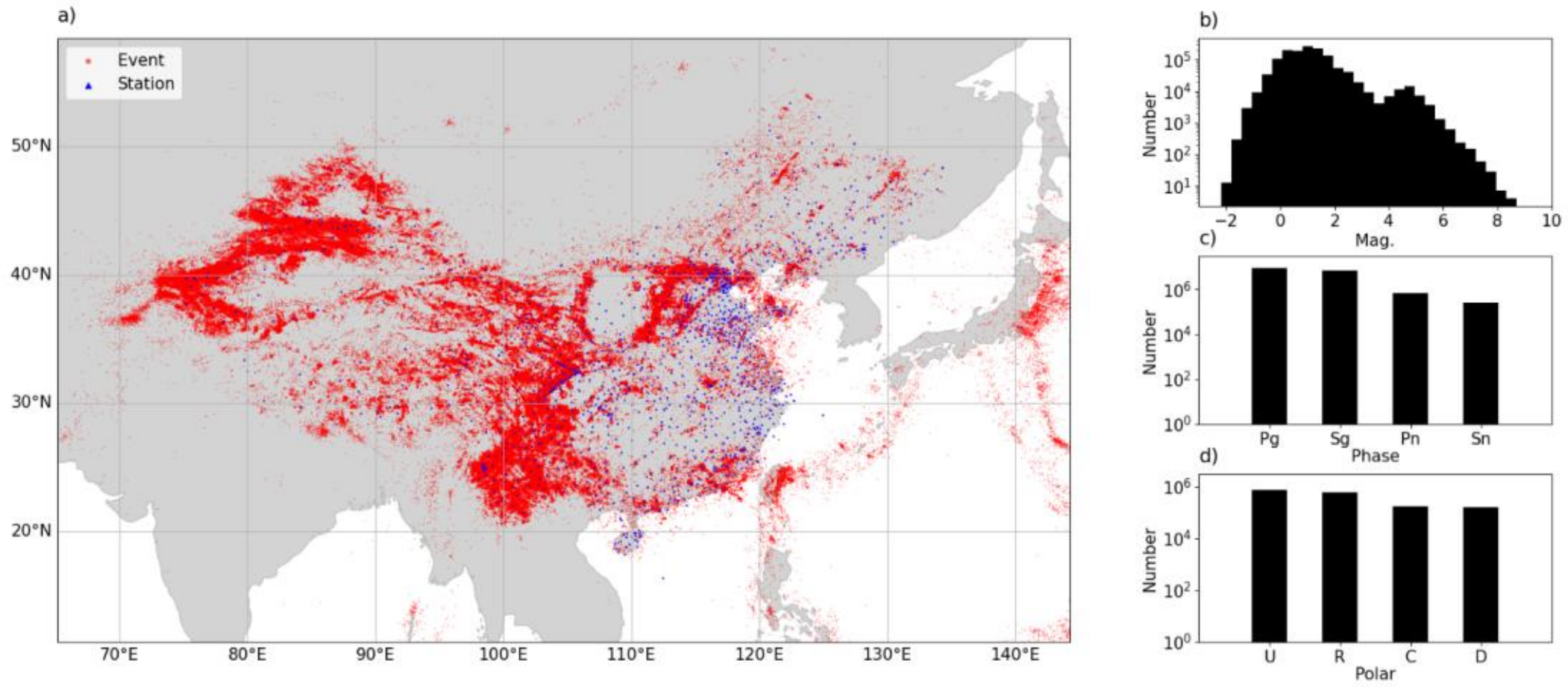

Figure 1 The training dataset.
(a) The location of the events and station; (b) the number of different phases; (c) the magnitude of different events.

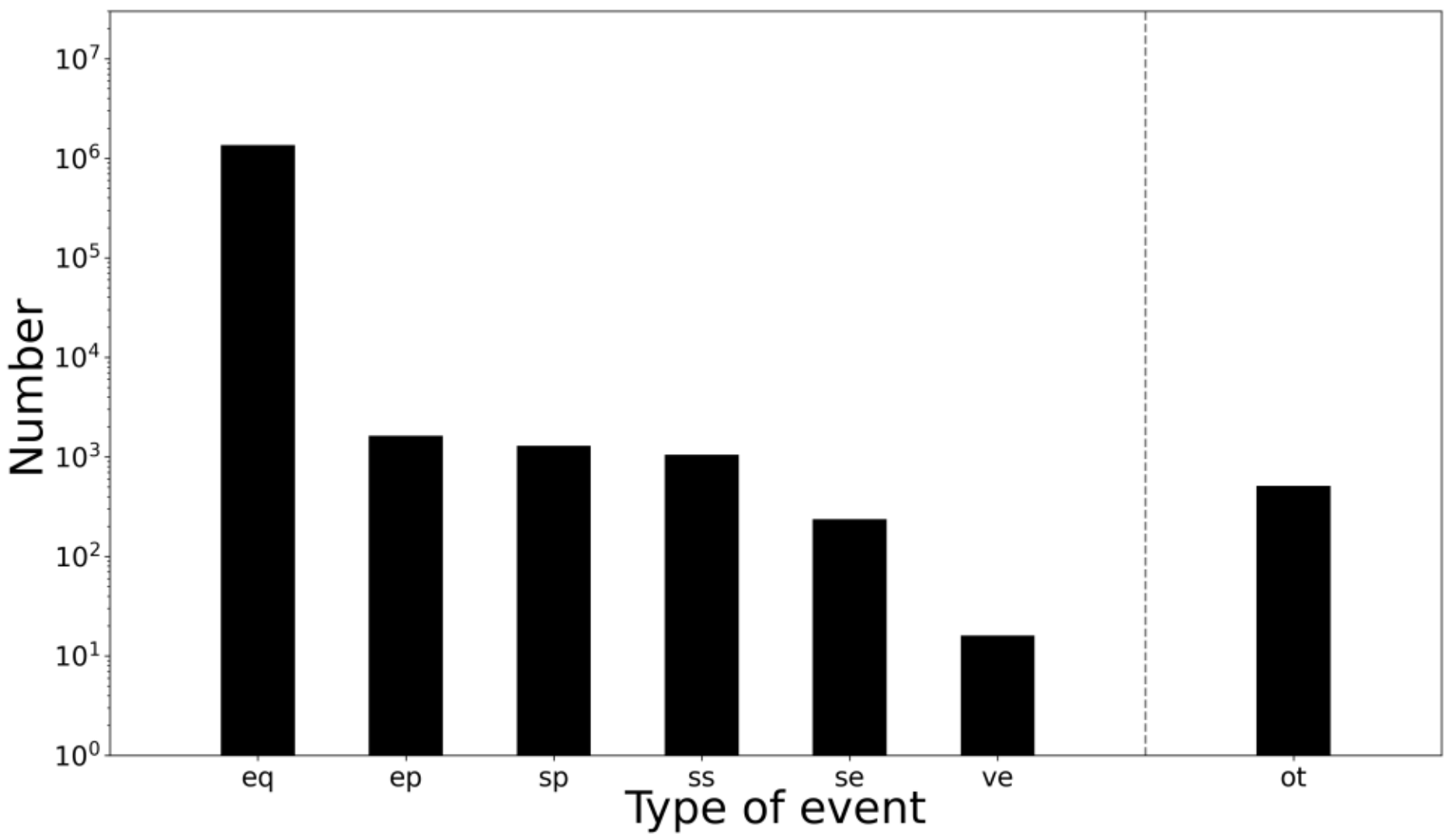

Figure 2 The number of different type of events.

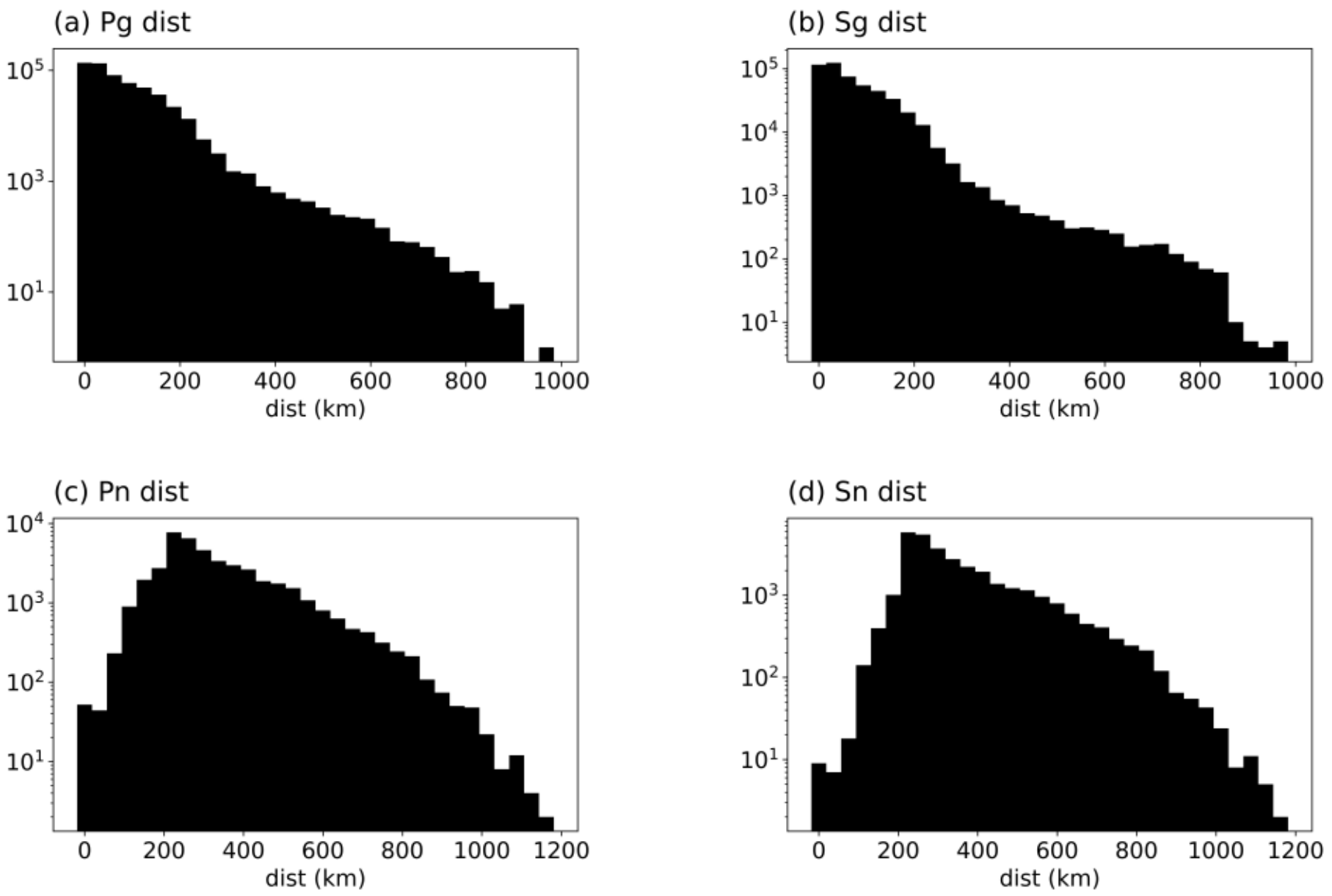

Figure 3 The statics of epicenter distance for Pg, Sg, Pn and Sn phases.

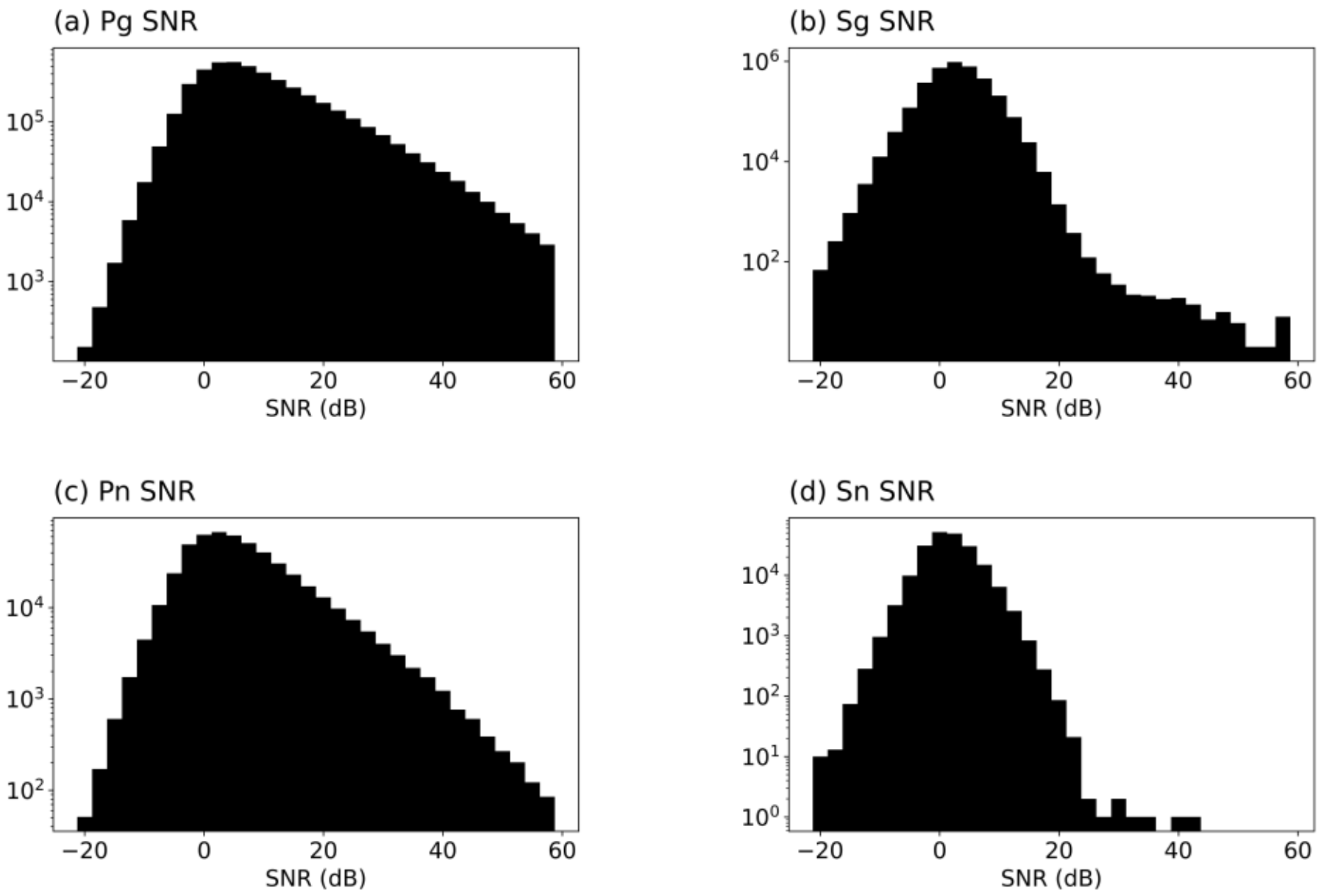

Figure 4 The distribution of different SNR.

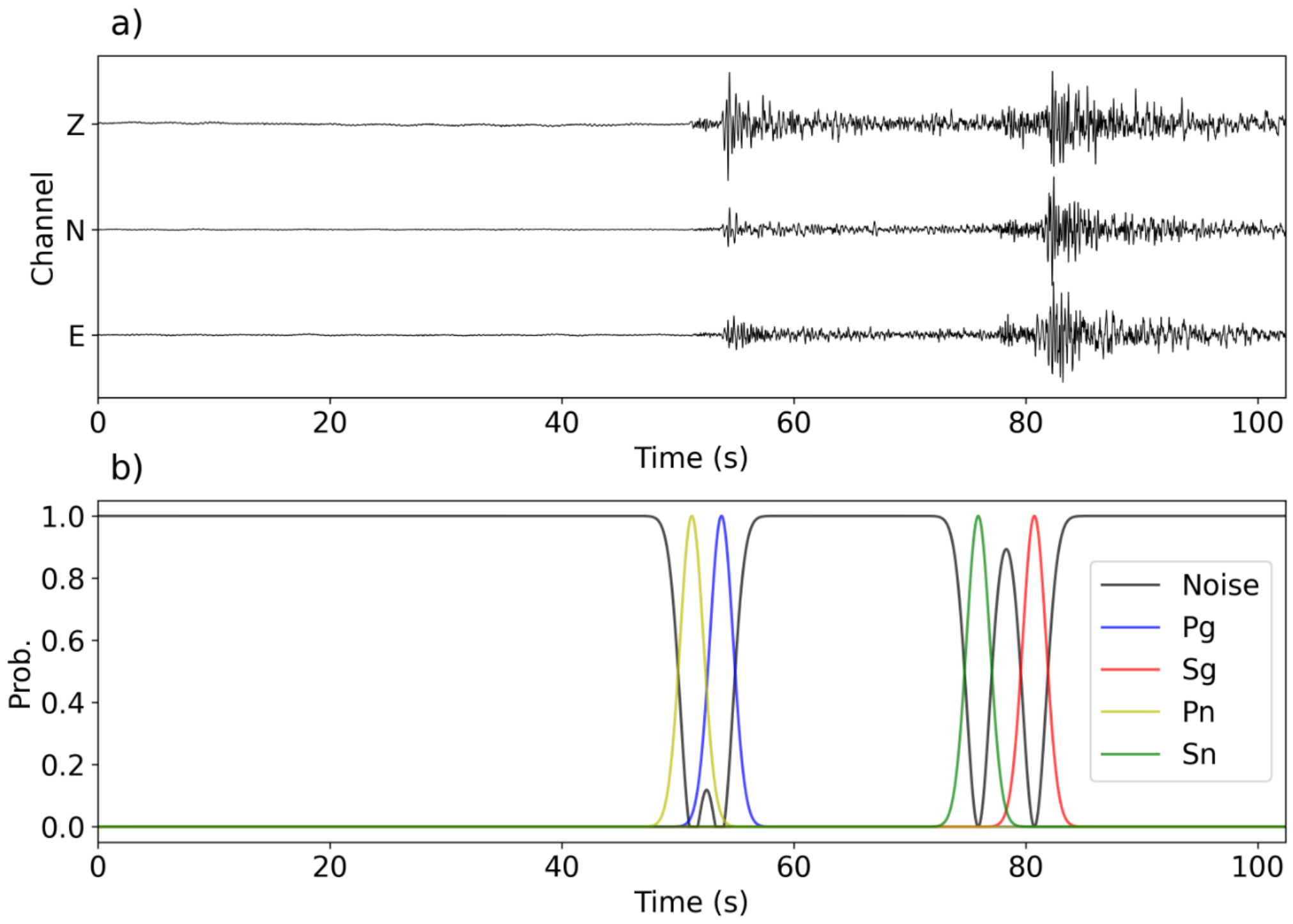

Figure 5 The input wavefrom (a) and the label of phase (b)

## 2.2 The architecture of our model

When extract multimodal information from seismic waveform data, a larger model capacity is required, which contains trainable parameters are necessary. When constructing our pre-trained network, we want to include as much manually labeled information as possible. Therefore, we used bi-directional transformer as backbone of our model. Additionally, to accommodate waveform input and output, we incorporated convolutional neural networks into our architecture, as illustrated in Figure 6.

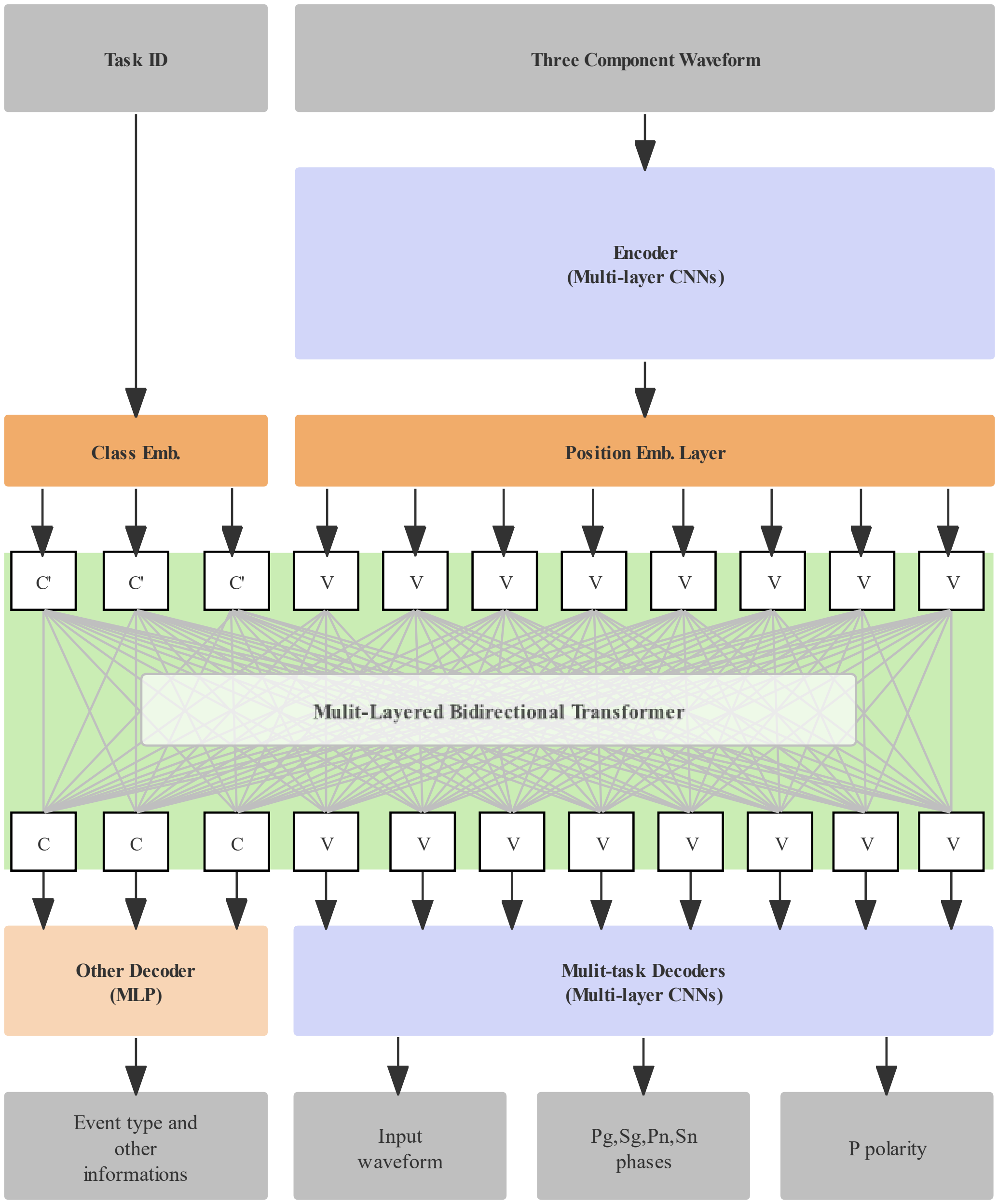

Figure 6 The architecture of our model.

To enhance the extraction of seismic waveform features, we take raw three-component seismic waveforms as input. The length of the seismic waveform is 10240 sampling points, which is 102.4s waveform. The input data is $x \in \mathbb{R}^{3\times N}$, that $N = 10240$ or can be set to any other length. As the pre-trained model will output the estimation of location of an event and the P polarity, it is necessary to arrange three component data in E, N, Z order. We normalize three component's data by its maximum value without any additional preprocessing.

There are mainly five components in our model: waveform encoder, task embedding layers, position embedding layers, bi-directional transformer layers and multi-task decoders. Counting that CNN are suitable for waveform, we employ multi-layer CNNs to construct waveform encoder for processing the raw waveforms, obtaining waveform features vector $u_{wave} \in \mathbb{R}^{C\times T}$, $C$ is the length of vector and $T$ is time step of vector. To add positional information into $u$ we incorporate a position encoder built with bidirectional recurrent neural networks to encode position information. There are task embedding layers to output different type of task vector $u_{task} \in \mathbb{R}^{C\times M}$, where $M$ is the number of desired tasks. $M$ is set to be 3 and it is used to event classification, single station location estimation. The waveform feature and the task vector are concatenated, and the vector will be $u = concat[u_{wave}, u_{task}] \in \mathbb{R}^{C\times L}, L = T + M$. Atop this, we add multi-layer transformer as backbone of our model, like the bidirectional nature of the BERT model, which allows for a comprehensive consideration of all waveform features and outputs feature vectors $v \in \mathbb{R}^{C\times L}$.

As our model need to output multiple information such as phases, earthquake types, we introduce multi-task decoding, which provides additional features for subsequent analysis. The decoders are divided into Type A and Type B, both of which take the feature vector $v$ as input, albeit with slight differences in their configurations. Type A is based on a fully-connected layer (see Figure 2c, where the final feature number K is specified by different tasks) and takes $v_{task} \in \mathbb{R}^{C\times M}$ as input. The out of Type B can be $o_{typeA} \in \mathbb{R}^{K}$. Type B employs an up-sampling layer (using interpolation for upsampling) combined with convolutional neural network layers (collectively referred to as transposed convolution) to construct the basis (see Figure 2b, where the final filter number K is specified by different tasks), and takes $v_{wave} \in \mathbb{R}^{C\times T}$ as input. The output length of type B decoder matches the waveform length $o_{typeB} \in \mathbb{R}^{K\times N}$. Different decoders can accomplish various tasks. Type B decoder is used to phase picking, P polarization estimation and waveform reconstruction.

During the pre-training process, we incorporated four main decoders. The first decoder (Type A), corresponding to the multi-task encoding vector, outputs earthquake types; there are seven types of earthquakes specified with K=7. The earthquake type decoder outputs from the task code 0 vector using a fully connected network, ultimately outputting features for seven types, corresponding to the eight earthquake types in the CSNCD (natural earthquakes, explosions, landslides). However, it is important to note that the dataset is highly imbalanced, with natural earthquakes accounting for 99.5% of samples, suggesting a bias towards predicting earthquakes as natural. Here, we prefer to use local data for transfer learning to address this issue, detailed in Chapter 3. The seismic phase type decoder translates the general pre-trained feature to match the waveform length and predicts the phase type.

The second decoder (Type B) outputs phase types, with the output length matching the length of the seismic waveform. During training, we included four phase types with relatively abundant manually annotated data: Pg, Sg, Pn, and Sn, totaling five types including noise, specified with K=5. The third decoder (Type B) output types of P-wave initial motion directions, with two types specified, K=2. The initial motion decoder still uses point-to-point output, considering multiple seismic signals may be included in the decoding process. The fourth decoder (Type B) outputs the original seismic waveform, comprised of three channels, specified with K=3. It encodes more waveform information, and since the autoencoder structure outputs the waveform itself without label information, it is unsupervised learning.

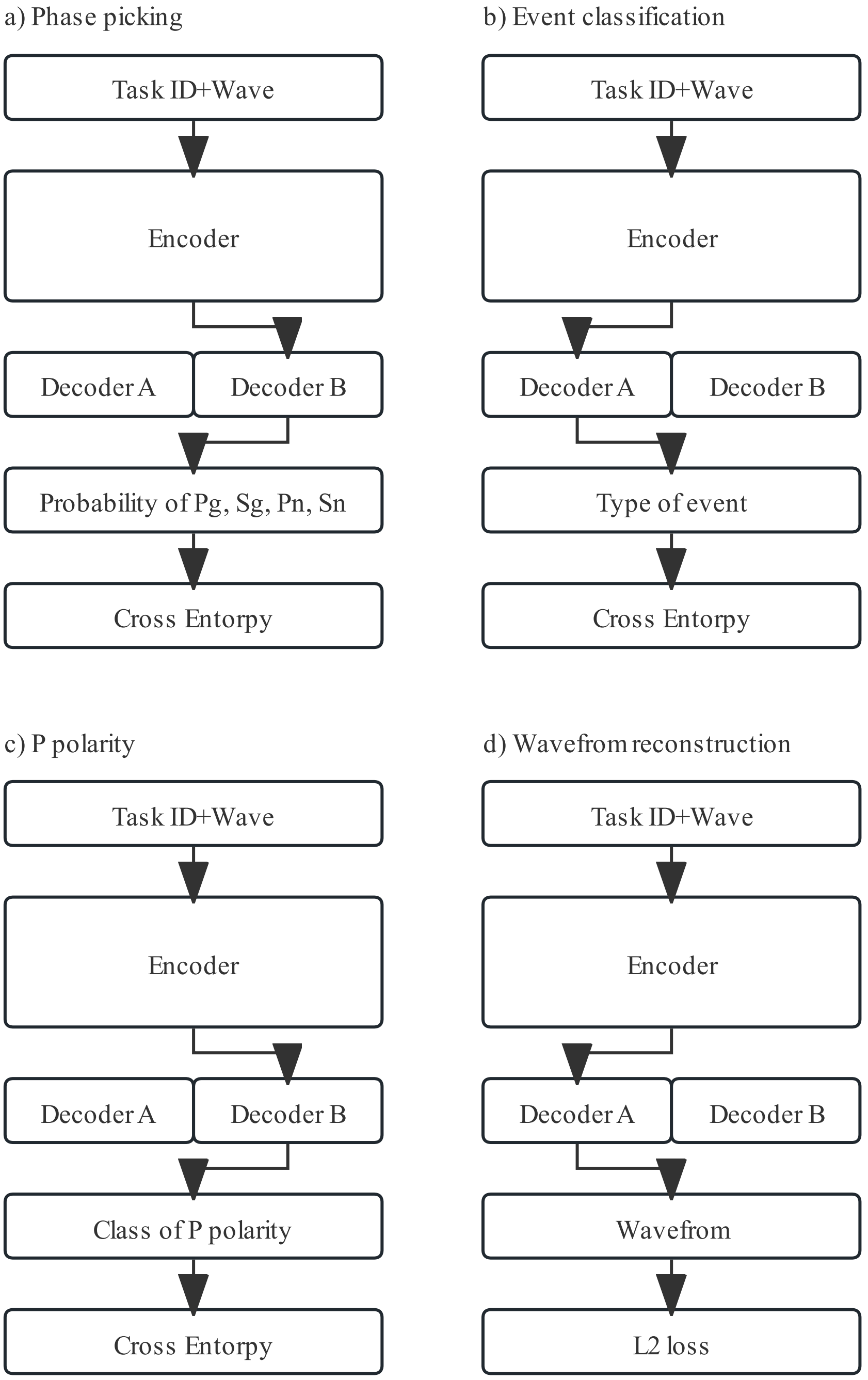

Figure 7 The flowchart of different tasks.

The diagram illustrates the computational processes of the four decoders. Additionally, we have designed four loss functions for different decoders:

$$loss = w_{type} \cdot loss_{type} + w_{phase} \cdot loss_{phase} + w_{polar} \cdot loss_{polar} + w_{wave} \cdot loss_{wave}$$

For the earthquake type decoder, the output $y^{type} \in \mathbb{R}^{K}$, where K represents the number of data samples and 7 corresponds to the seven earthquake types. The type of decoder uses cross-entropy as the loss function, without any weighting across different types. The loss function can be described as:

$$loss_{type} = \log\left(\frac{1}{p_c^{type}}\right)$$

$c$ is the manually labeled type of index. $p^{type}$ is the softmax from $y^{type}$.

In the phase decoder, each sampling point requires a prediction of its category $y^{phase} \in \mathbb{R}^{K\times N}, K = 5$, hence each point employs cross-entropy.

$$loss_{phase} = \sum_{i=1}^{N}\sum_{k=1}^{5} d_{i,k}^{phase} \log\left(\frac{1}{p_{i,k}^{phase}}\right)$$

$d^{phase}$ is the label of different type of phase, which is described by Yu et al. (2023). $p^{phase}$ is the softmax from $y^{phase}$.Similarly, the initial motion labels use cross-entropy, but these labels $d^{polar} \in \mathbb{R}^{2\times N}$ are only present for the P-wave sections, not across all phases and waveforms. Therefore, we apply a weight $w \in \mathbb{R}^{N}$ of 1 to the 0.5 seconds before and after the P-wave, and a weight of 0 elsewhere where P-waves are absent.

$$loss_{polar} = \sum_{i=1}^{N} w_i \sum_{k=1}^{5} d_{i,k}^{polar} \log\left(\frac{1}{p_{i,k}^{polar}}\right)$$

In the waveform decoder, constraints are used to generated data. So, we use MSE loss to between the output of decoder wave $y^{wave}$ and $x$. Here, L2 regularization is used in the loss function to constrain the range of the feature vector $v$. This part pertains to the variational autoencoder and represents unsupervised machine learning.

$$loss_{wave} = \sum_{i=1}^{N}\sum_{k=1}^{5}\left(x_{i,k} - y_{i,k}\right)^2 + L_2(v)$$

These loss functions are weighted by $w_{phase}, w_{polar}, w_{type}, w_{wave}$ that each of the losses has similar scale.

# 3. The test of our model

## 3.1 The accuracy of picking Pg、Sg、Pn、Sn phases

To evaluate the performance of our model in the task of phase picking, we conducted testing using samples within a 2000 km epicentral distance on test dataset. To ensure the objectivity and comprehensiveness of the evaluation, we used standard evaluation metrics, including precision (P), recall (R), F1-sore (F1), mean (μ) and standard deviation (σ) of time residuals between predicted and labeled arrival times. There are three types of picked samples: true positive samples (TP), false positive samples (FP) and false negative samples (NP). If the misfit between the pick and the manual pick was less than 1s, it was defined as a TP sample; otherwise, it was defined as a FP sample. If no TP picks were found for a manual pick, the sample was defined as an FN sample. Precision ($P = \frac{TP}{TP+FP}$) is the fraction of TP among all picks, and recall ($R = \frac{TP}{TP+FN}$) is the fraction of TP picks among all possible picks. The F1-score ($F1 = \frac{2PR}{P+R}$) is defined as the harmonic mean of precision and recall, and it provides a more detailed and balanced evaluation than precision or recall alone.

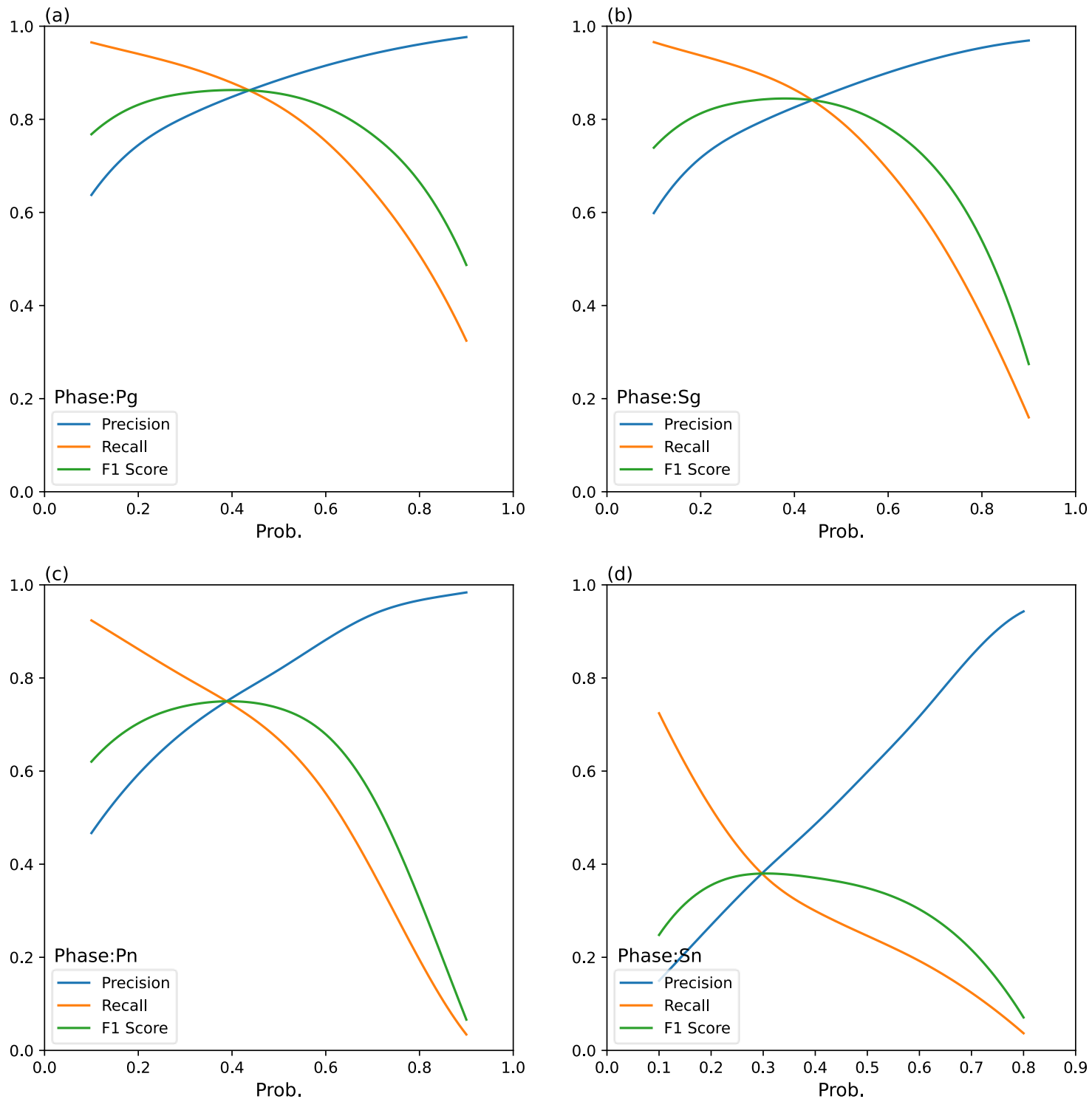

Figure 8 P-R Curve

The picking probability threshold is a crucial factor affecting the precision, recall, and F1 score of a model. As demonstrated in Figure 8, these metrics vary with changes in the set threshold. Typically, as the threshold increases, the recall decreases while the precision increases. This is because a higher threshold requires stronger evidence for a positive prediction, reducing the number of false positives but potentially missing some true positives. For different seismic phases, there should be corresponding optimal thresholds to maximize the F1 score, which balances precision and recall. According to Figure 8, the phases Pg, Sg, Pn, and Sn show higher F1 scores at thresholds of 0.43, 0.43, 0.40, and 0.3, respectively. These thresholds reflect the point where each phase achieves the best trade-off between missing real events (lower recall) and incorrectly identifying events (lower precision). For subsequent experiments in phase detection, we use these probability thresholds for each phase.

Table 1. The Statistical Metrics of Model. The mean and standard deviation are calculated with the picks, whose picking error between manual and auto are smaller than 1.5 seconds.

| Phase | $\mu(ms)$ | $\sigma(ms)$ | P | R | $F_1$ |
|---|---|---|---|---|---|
| Pg | -65.52 | 383.26 | 0.860 | 0.866 | 0.863 |
| Sg | -33.31 | 463.34 | 0.838 | 0.846 | 0.842 |
| Pn | -287.23 | 738.94 | 0.757 | 0.743 | 0.750 |
| Sn | -55.77 | 1386.36 | 0.383 | 0.377 | 0.380 |

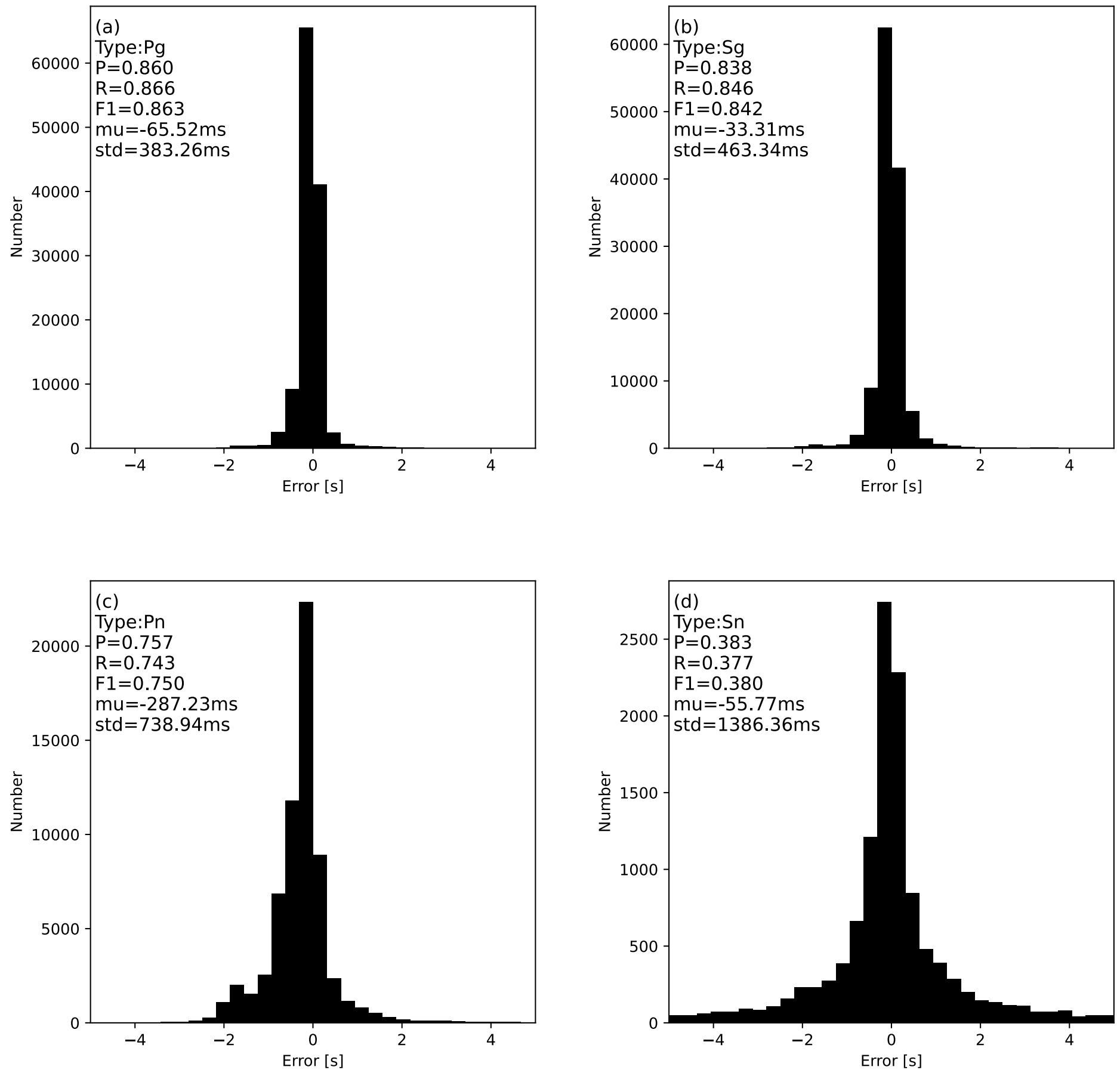

Figure 9 The accuracy of different phases on test dataset.

Under optimal probability thresholds, the performance of the model for different types of seismic phases in terms of precision, recall, and F1 score is outlined in displays the distribution of time residuals between the model-picked and manually annotated phase arrivals Figure 9. From Figure 9 and Figure 8Table 1, the model exhibits higher recognition accuracy for Pg and Sg phases, which are more abundant in the dataset, allowing the model to learn their characteristics effectively.

The Pg phase achieved the highest precision, recall, and F1 score at 0.860, 0.866, and 0.863, respectively. The histogram statistics show that the arrival time error distribution for the Pg phase is centered around -65.52 ms, with standard deviation 383.26 ms. For the Sg phase, the precision, recall, F1 score and standard deviation are lower than Pg. However, the mean of the errors is slightly higher than Pg phase. This is consistent with the fact that S-waves are more susceptible to interference from P-waves and are thus more challenging to accurately pinpoint the arrival time.

In contrast, the recognition accuracy for Pn and Sn phases is lower, but the accuracy for Pn phases is higher than for Sn phases. The precision, recall, and F1 score for the Pn phase are 0.757, 0.743, and 0.750, respectively, indicating that the model has higher accuracy in detecting the Pn phase. The mean error for Pn is -287.23 ms with a standard deviation of 738.94 ms. As seen from Figure 9, the error distribution is relatively wide and skewed towards negative values. This means that Pn waves might be influenced by interference from Pg waves. In the waveform data, Pn and Pg waves may arrive within a similar time window, especially at shorter epicentral distances, where this overlap increases the difficulty of differentiation, potentially causing the model to confuse these two phases, thus affecting the accuracy and reliability of Pn phase detection.The Sn phase exhibits the lowest accuracy among all phase detections, with the histogram showing the widest error distribution. The mean error for Sn is -55.77 ms, with a standard deviation of 1386.36 ms. This could be due to the number of Sn phase samples is the smallest in the dataset, resulting in insufficient data samples during the training process for the model to adequately learn its waveform characteristics. Additionally, the Sn phase signals are generally weaker compared to Pg, Sg, or Pn phases and are more susceptible to noise interference, which increases the difficulty of detection. Nonetheless, the model achieves a precision of 0.389, a recall of 0.377, and an F1 score of 0.380 for Sn phase detection, indicating that the model still possesses basic performance capabilities for detecting the Sn phase.

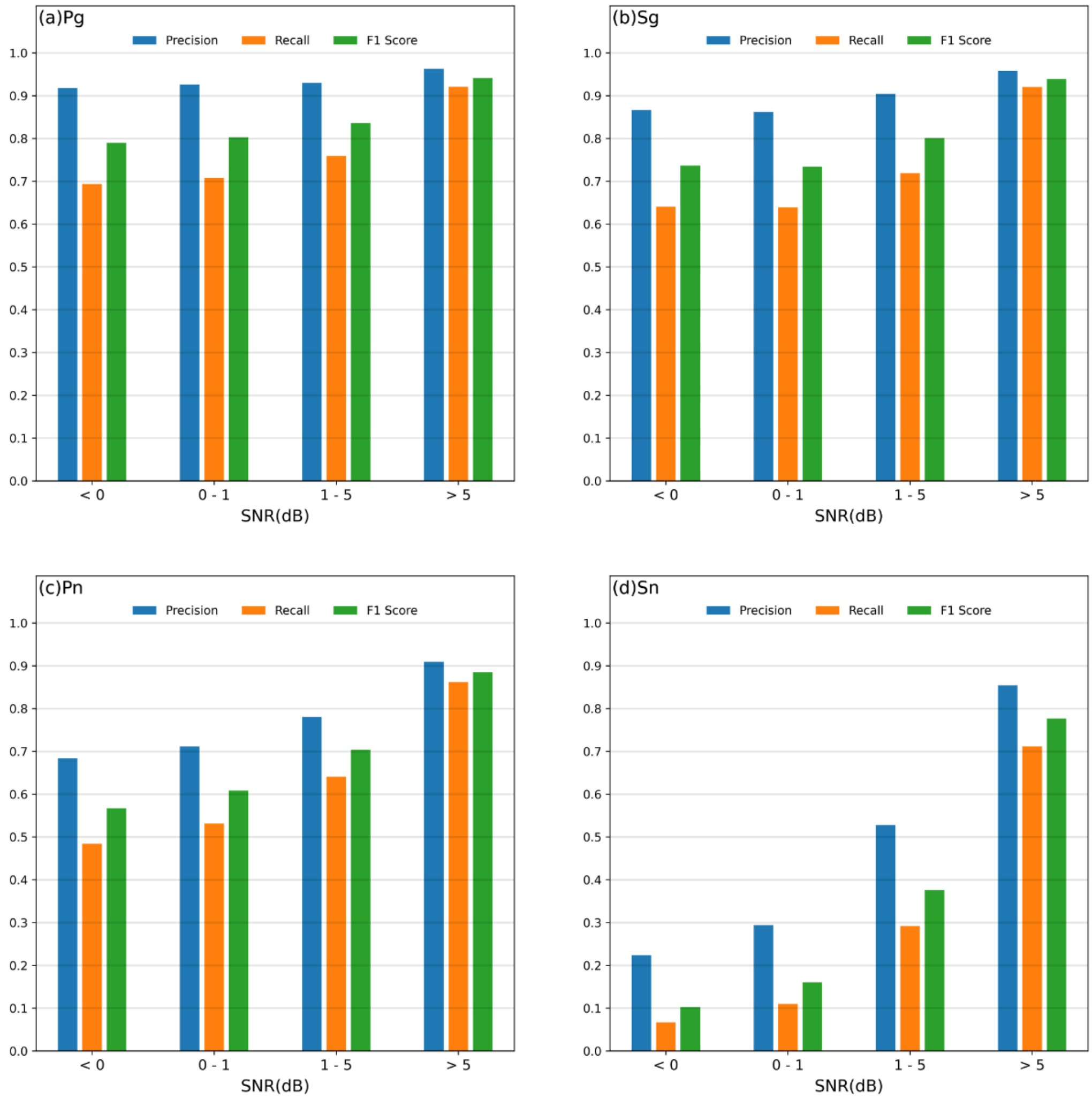

Figure 10 The accuracy of Pg, Sg, Pn and Sn phases on different SNR.

Many studies have shown that the signal-to-noise ratio (SNR) is an important factor affecting the accuracy of model pickups. SNR reflects the relative strength of the signal against background noise, and for seismic signal detection, a higher SNR means the signal is clearer, which facilitates the model's accurate phase identification. Figure 10 displays the model's performance across different SNR categories, where precision, recall, and F1 score are calculated for each SNR category. It can be observed that as SNR increases, the model's performance improves consistently. Specifically, with the increase in SNR, there is a stable increase in precision, recall and F1 score, indicating that the model is more likely to accurately identify seismic phases under clearer signal conditions. We can see that the picking accuracy of Pn and Sn are more easily affected by Pg and Sg phases. As the SNR increases, the model's performance also improves. This enhancement is particularly noticeable for the Sn and Pn phases; when the SNR is greater than 5dB, both achieve higher accuracy. Even under low SNR conditions, the model maintains high precision for Pg phase, while the change in recall is more pronounced, further indicating that under low SNR conditions, the increase in noise interference makes it more difficult to distinguish between seismic signals and noise.

## 3.2 Test on P polar classification

To thoroughly assess the performance of our model on the initial motion determination task, we conducted detailed analyses using the CSNCD dataset's test set. Given that the dataset includes annotations for the clarity of polarity, we conducted two independent tests. The first test dataset marked with I clarity (16,590 samples), which represents the polarity is clear. The second dataset includes all clarity levels (48,400 samples), which encompasses I clarity, M clarity and E clarity (E means that the polarity is not clear and M means that the clarity is not marked). I, M and E are manually marked.

We visualized the model's performance across different polarity categories using confusion matrices. As shown in Figure 11a, when testing with the highest clarity "I" dataset, the model demonstrated high performance, achieving an accuracy of 88.54% for upward and 88.36% for downward. This result indicates that under high clarity conditions, the model can precisely determine the direction of earthquake initial motions. As illustrated Figure 11b, when we combined data of all clarity for testing, there was a slight decline in performance. The model achieved an accuracy of 83.38% for upward and 80.28% for downward.

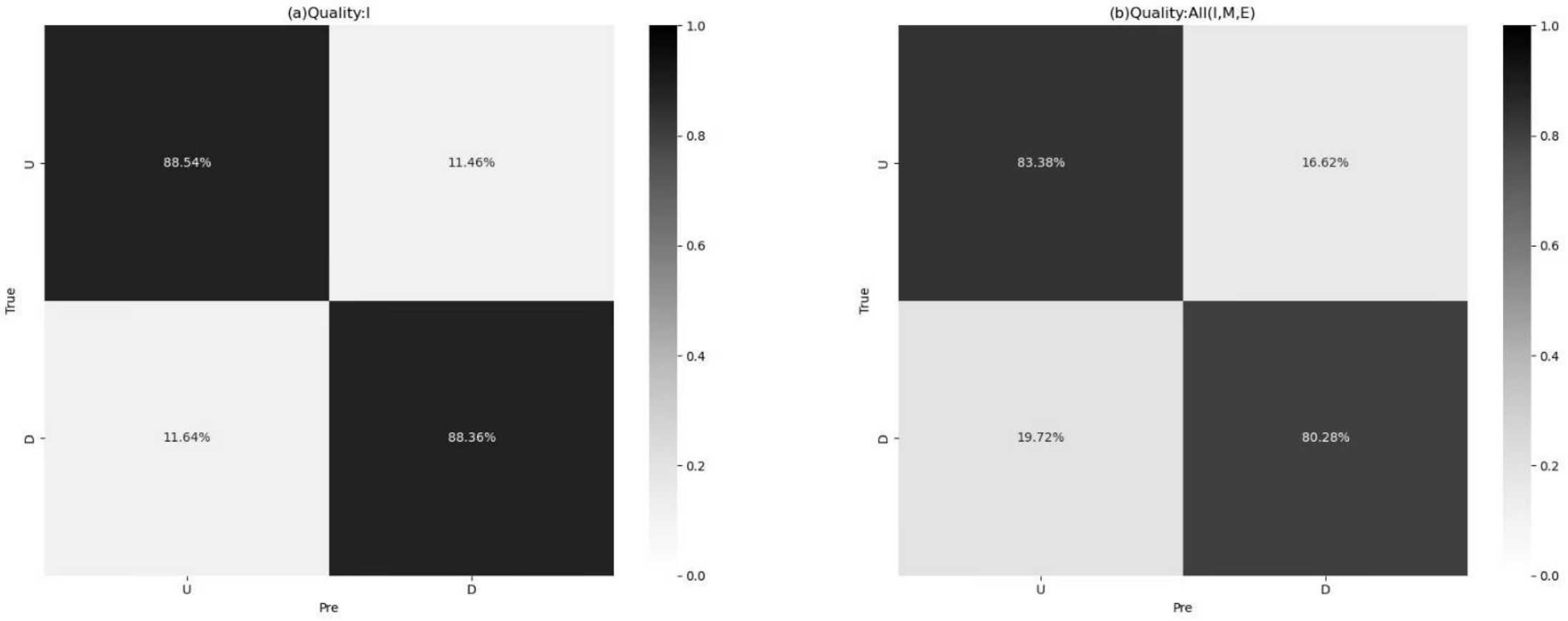

Figure 11 The confusion matrix of P polarity. Test on I clarity (left) and all clarity (right).

## 3.3 Earthquake classification

We selected data from the Inner Mongolia region for our classification test, comprising a total of 419 events, of which 251 were used for training and 168 for testing. In this region, there are three main types of seismic events: earthquakes (EQ, 103 events), explosions (EP, 244 events), and subsides (SS, 72 events). The explosions and subsides are more concentrated, whereas the earthquakes are dispersed throughout the testing area (Figure 12a). The magnitudes of the earthquakes are concentrated between 1.5 and 3.5, with a lack of records for smaller earthquakes in this region (Figure 12b red). The depth of explosions and subsides are concentrated near the surface, with a focal depth close to 0km, while earthquakes are distributed at depths between 0-30km (Figure 12c). During the training process, we limited the maximum epicentral distance to 200km, with most non-natural earthquake events occurring within 75km (Figure 12d).

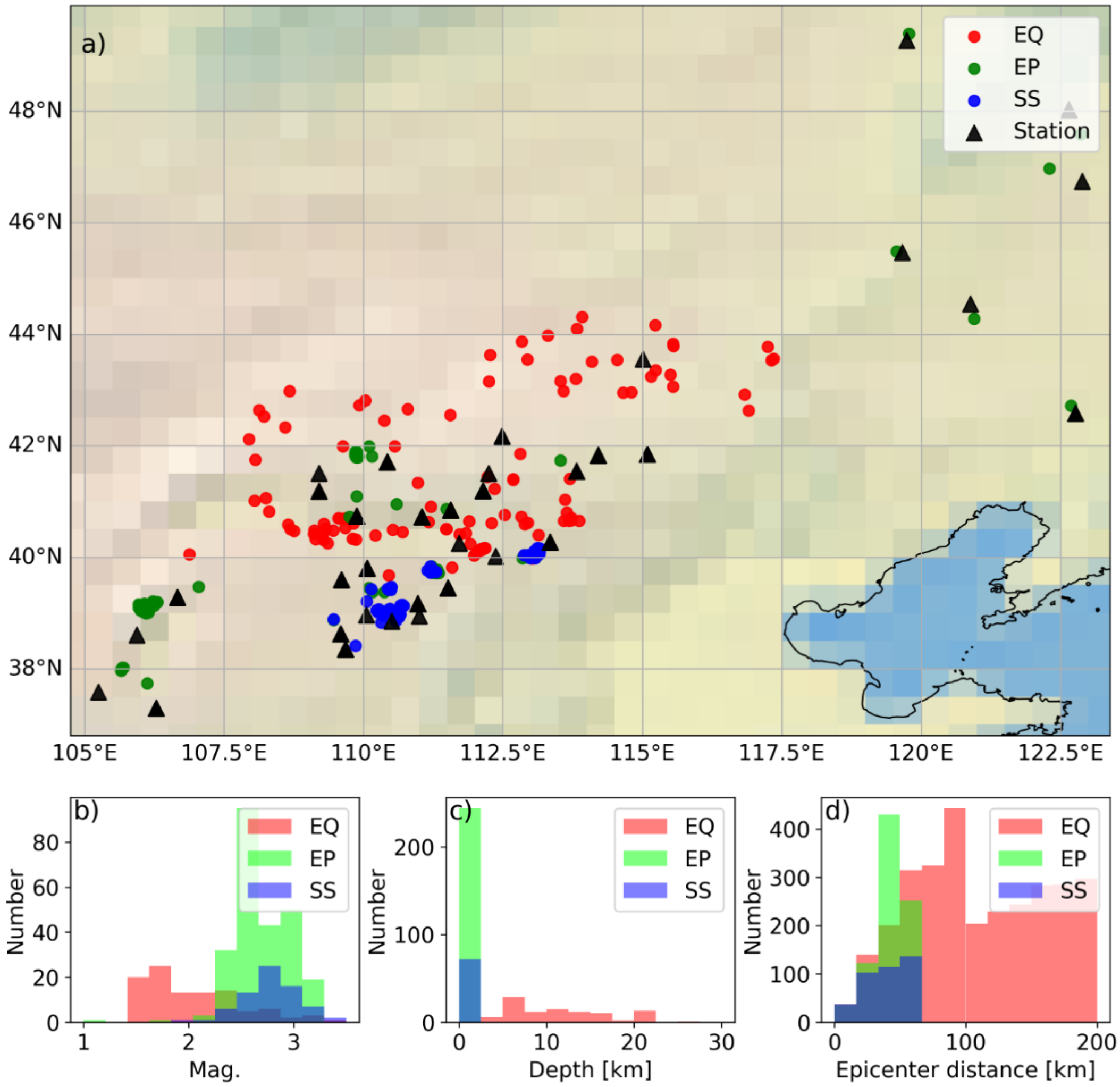

Figure 12 The classification dataset.
a) The location of different type of event and station; b) The distribution of magnitude of different type; c) The distribution of depth; d) The distribution of epicenter distance.

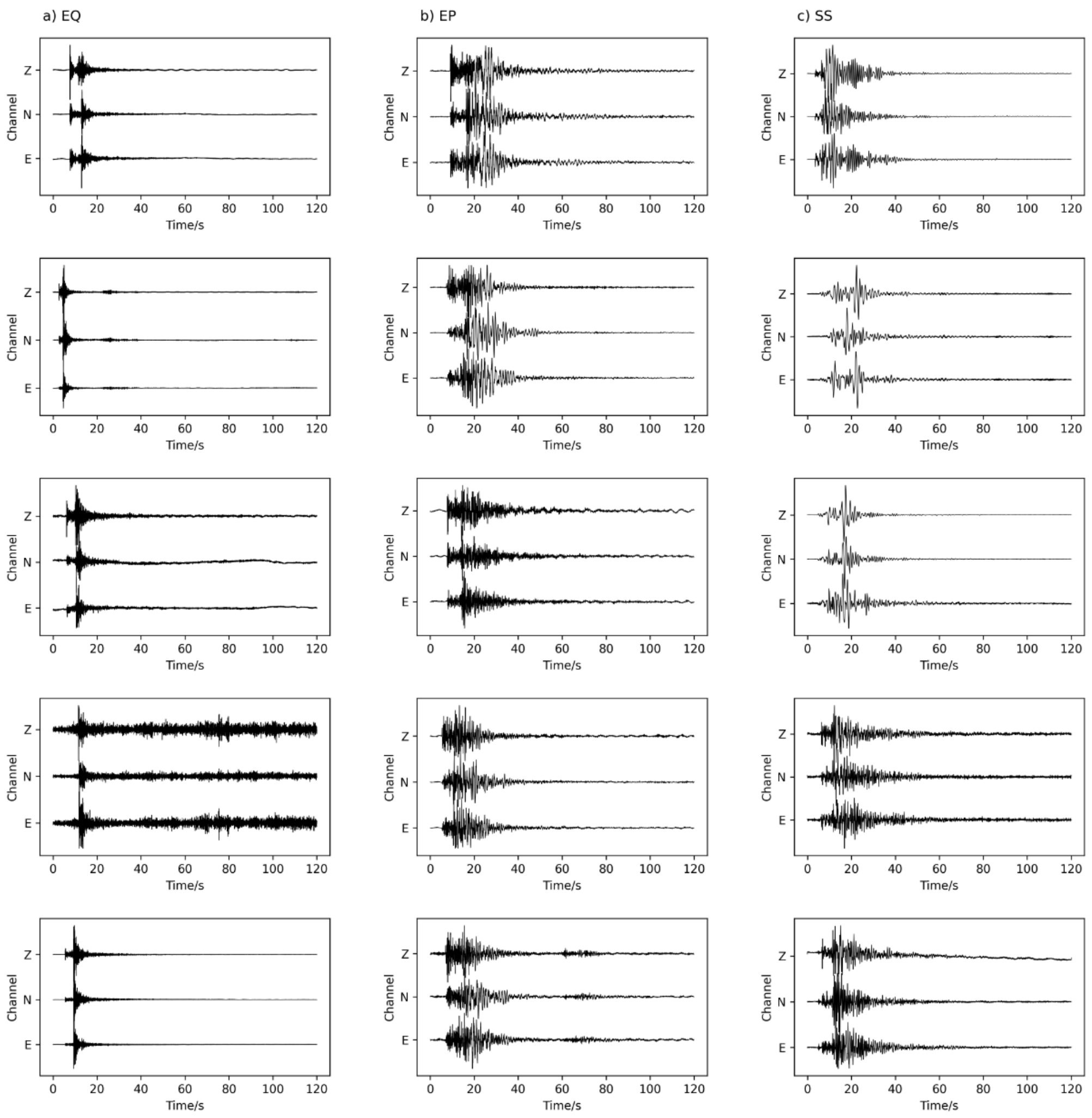

Figure 13 The waveform of different type of event

It can be seen from Figure 13 that earthquake events have a higher P/S amplitude ratio and a richer frequency component (Figure 13a). In contrast, explosion events have a larger P/S amplitude ratio compared to natural earthquakes and collapses, but the S-waves are less distinct. Collapse events have a smaller P/S amplitude ratio and more low-frequency components.

We use two models to test the performance of our model: the first is trained directly without transfer learning; the second undergoes transfer learning using a generic pre-trained model. Both models are iterated 501 times with a fixed learning rate of 1e-5. Strictly speaking, our earthquake classification is not an entirely new task, as there were classification labels during the pre-training process. However, due to extreme sample imbalance in direct training (natural earthquakes comprise 99.5% of the data), the classification output is not available. Therefore, it is necessary to use local data for transfer learning.

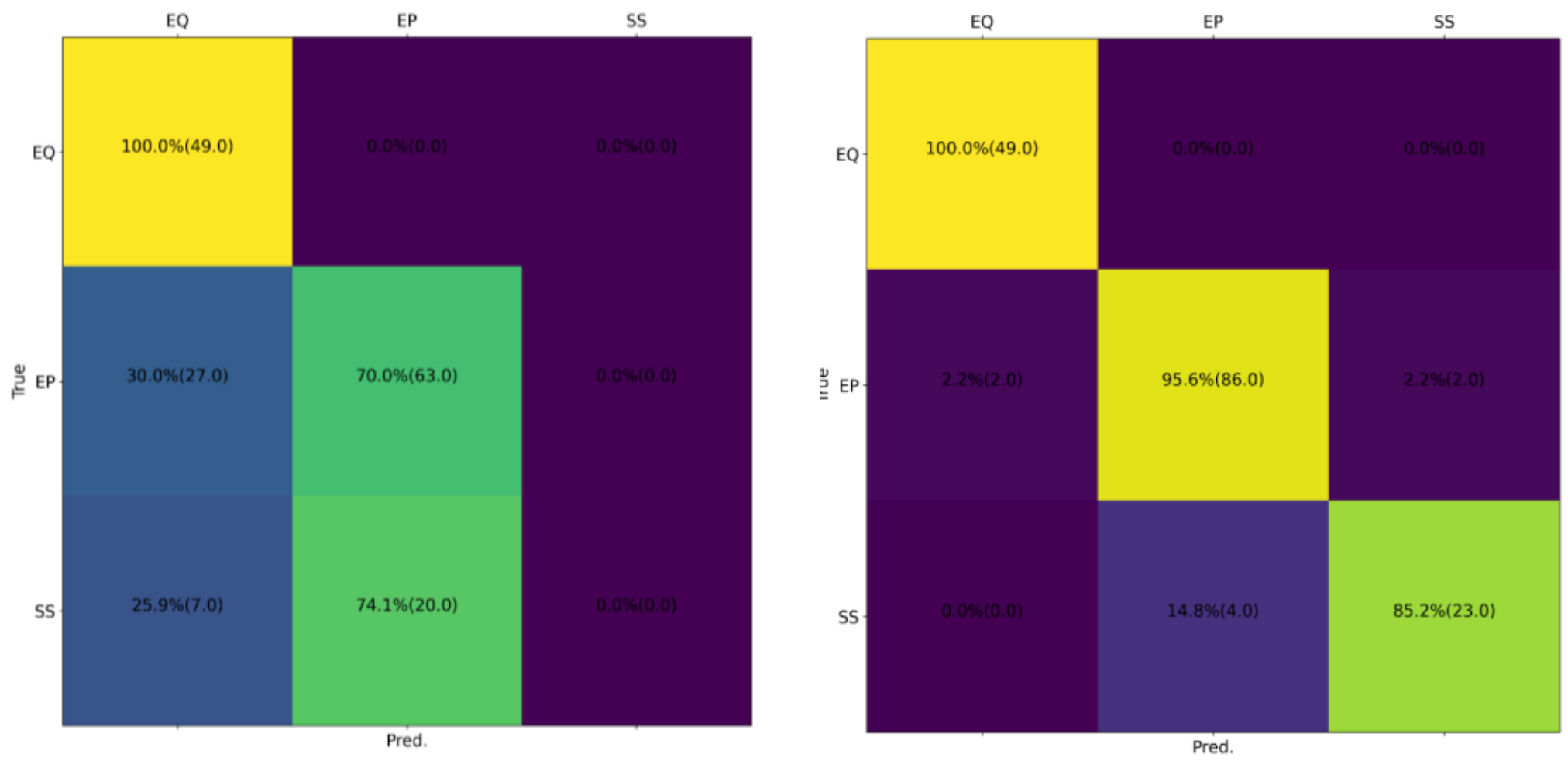

Figure 14 The confusion matrix of different models
Newly trained model (left) and transfer learning model (right)

During training, only the decoder of event classification is trainable. A single earthquake can be detected by different stations, and currently, our model is a single-station model. Therefore, after classifying at individual stations, we tally which class has the most predictions for the same earthquake to determine the actual type of earthquake. In this scenario, the accuracy of the model trained from scratch is 67.5%. After transfer learning, the classification accuracy improves to 95.2%. It is observed that before transfer learning, all collapse events were classified as explosions, due to the predominance of explosion events and the impact of sample imbalance, resulting in all collapses being misidentified as explosions (Figure 14). Additionally, many events were inclined to be identified as natural earthquakes, due to the uneven distribution of stations recording natural events. However, after transfer learning, there was a significant improvement in accuracy; natural earthquakes were successfully identified, while collapse events were more frequently recognized as explosions, suggesting that some collapse events might share similarities with explosions based on their waveforms.

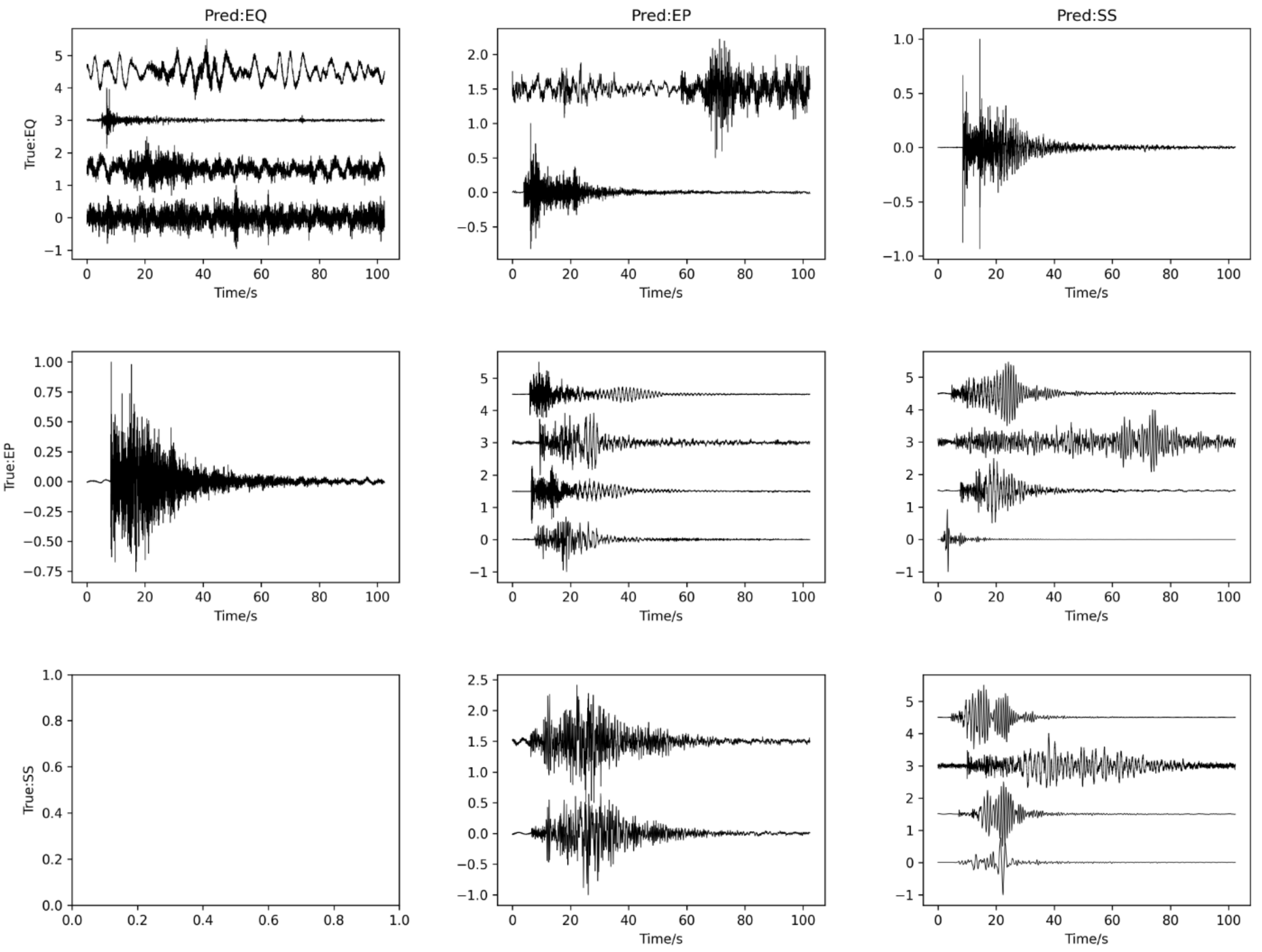

Figure 15 The confusion matrix of waveform for different type of event (Z component)

In Figure 15 we collect four waveforms for each classification result. However, there is no false samples such as SS incorrected classified to EQ, so we do not have sample to plot. It can be observed that explosions misidentified as natural earthquakes have similar waveform characteristics, specifically, a smaller P/S amplitude ratio. Similarly, natural earthquakes predicted as explosions also share this trait, i.e., a larger P/S amplitude ratio or very close arrival times between P and S waves. The misidentified waveforms of collapses that were recognized as explosions exhibit distinct collapse characteristics, such as a prevalence of low-frequency components and a small P/S amplitude ratio. However, the neural network failed to correctly identify these features, indicating that purely waveform-based classification lacks sensitivity to certain characteristics like frequency. Furthermore, the training data also contain waveforms like those of explosions, which is another reason contributing to the misclassification by the model.

# 4. Discussion

## 4.1 Phase picking

In the testing of phase detection, the model achieved an F1 score of 0.863 for Pg phase and 0.842 for Sg phase (F1 score combines precision and recall into a harmonic mean). The accuracy for Pn phase was relatively lower at 0.750, with a broader error distribution. The test accuracy for Sn phase detection is lowest at 0.380. This may be due to the Sn phase being weaker compared to other phases, making it relatively difficult to distinguish in seismic waveforms, especially under low signal-to-noise ratio conditions where the Sn phase's waveform characteristics are not pronounced enough. The model's inability to accurately learn the features of Sn phase can lead to misclassifications as noise during testing, thus reducing detection accuracy. Additionally, the dataset contains fewer instances of Sn phase compared to Pg and Sg phases. This discrepancy in sample sizes may have led to insufficient learning of Sn phase features during training, thereby affecting its generalization capabilities on test data.

We also test the picking efficient on continuous data. Currently, many deep neural networks used for phase detection, such as PhaseNet, EQTransformer, and LPPN, are retrained on manually annotated datasets. However, the training data used (all three utilize the STEAD dataset) is limited. The maximum length of STEAD dataset is 60 seconds. That means that there are limited noise data before the Pg phase. However, the CSNCD dataset has at least 100 seconds waveform. That may affect the picking result of picking model. We selected 21/05/2021 and 22/05/2022 275 continuous waveforms, contains Ms6.4 YangBi main earthquake and after shock, to test our model and PhaseNet. The statics info is shown in Table 2.

Table 2. The accuracy between PhaseNet and our model. (We define a TP sample when the error between auto picked manual labeled time smaller than 1.0 seconds)

| Model Name | Phase | Recall | Mean of error (s) | standard deviation (s) | Total picks on continuous data. |
|---|---|---|---|---|---|
| PhaseNet | Pg | 0.871 | -0.031 | 0.188 | 919,281 |
| | Sg | 0.857 | -0.090 | 0.238 | 288,721 |
| Our model | Pg | 0.924 | -0.024 | 0.186 | 260346 |
| | Sg | 0.907 | -0.403 | 0.256 | 234009 |
| | Pn | 0.650 | -0.036 | 0.264 | 7117 |

We can see from the table that the recall of our model, on Pg and Sg phases, are 5% higher than PhaseNet. That means that our model can detected more phases. More training samples may make our model recognize more complex waveform, which make our model picking more phases.

More notably, our model detects fewer phases in continuous data at higher recall. That means that our model many greatly reduce false detection. And our model detects a more balanced amount of Pg and Sg, which is favorable for phase associations. As our model has longer input length of waveform, there is more noise data to reduce the false picking.

## 4.2 P polarity classification

There may be many Pg phases in 102.4 seconds time window. So, we used a point-to-point output to predict the P polarity. We build an independent P polarity classification model. The test accuracy is shown in Table 3

Table 3. The accuracy of independent P polarity model.

| Statics | value |
|---|---|
| Recall | 0.974 |
| Precision | 0.985 |
| F1 Score | 0.980 |

We can see that the independent model has higher accuracy than our model. The models have multi-task output, and the other output may affect the loss function during our network. We need to train the model with solely P polarity to get a higher precision.

## 4.3 Earthquake classification

The purpose of transfer learning is to extract waveform features from waveforms, which are used for various analyses and reduce the dependency on training data. Through seismic classification and teleseismic analysis, our general pre-trained model has initially achieved the above goals.

In seismic classification work, we can obtain a network model for seismic classification using a small amount of data (251 seismic events) through transfer learning. In this process, the classification effect of natural earthquakes is good, but the accuracy is low in explosion and collapse analysis. This may be because the neural network did not sufficiently learn waveform features, including spectrum and P/S amplitude ratio, which are important for seismic classification during pre-training. It is necessary to strengthen the training of the unsupervised waveform feature extraction part in the future. We found that more information such as source depth is equally important for seismic classification. For example, if earthquakes with a source depth of less than 1 km are classified as non-natural earthquakes and others as natural earthquakes, the accuracy of this classifier will reach 87%. However, a single model is difficult to effectively constrain earthquake depth, requiring multiple analysis models for processing.

# 5. How many parameters are sufficient

Yu et al. (2022) shows that the more parameters in LPPN the higher precision a model will get. However, Yu et al. (2023) shows that UNet++ has larger amounts of parameters with lower precision than EQTransformer and BRNN model. That shows that RNN/Transformer model has higher accuracy in analyzing seismic waveform. The BRNN and EQTransformer models have 0.45 and 0.44 million training parameters respectively. We built a P polar classification model with 0.96 million parameters and has 97% accuracy. That means that for a single task model, used to process single station waveform, less than 1.0 million parameters are enough. However, we need more parameters to process multiple information.

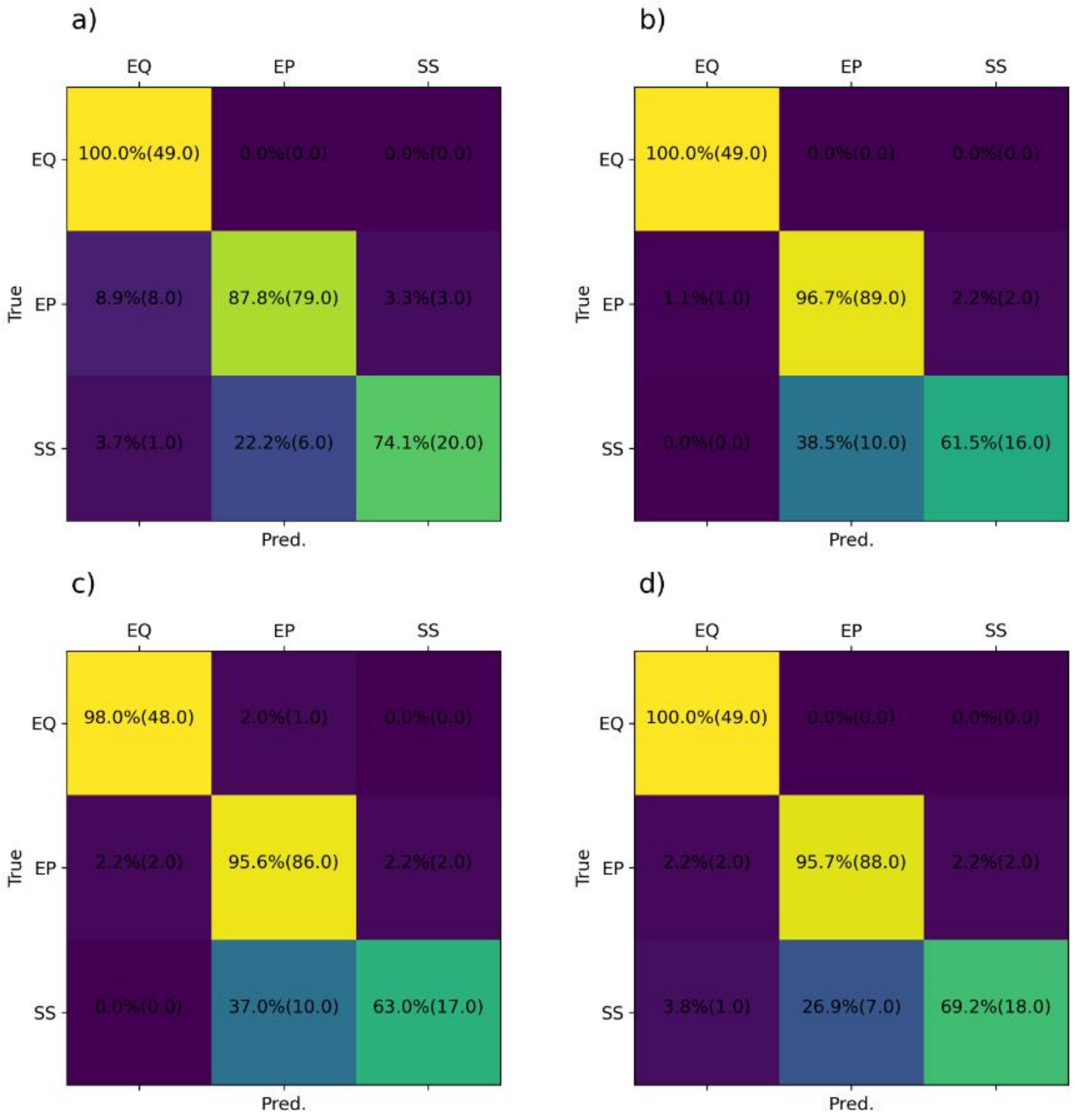

Figure 16 The training on a) model A, b) model B, c) model C and d) model D.

Table 4. The number of different model and test accuracy on event classification.

| Model name | Number of parameters (million) | Accuracy on event classification |
|---|---|---|
| Model A | 0.520 | 0.892 |
| Model B | 8.57 | 0.922 |
| Model C | 0.520 | 0.910 |
| Model D | 77.1 | 0.928 |

We built four models to further test the effect number of parameters. The accuracy is shown in Figure 16 and Table 4. Model A is a transfer learning model with 0.520 million parameters, whose classification decoder is trainable. Model B same to model A and has 8.57 million parameters. Model C share the same structure with model A and all the 0.520 million parameters are trainable. Model D has 77.1 trainable parameters and the transformer layers in 2.2Figure 6 are replaced by LSTM. We can see that the models with more parameters have achieve higher accuracy. Model C is a transformer learning model, and all the parameters are trainable. It has higher accuracy than model A and the accuracy on training dataset can be 100%. We can conclude that 0.51 parameters can be used for event classification. However, it is not sufficient to process multiple output. A newly trained model or all the parameters are trainable make the model suffer higher risks of overfitting. Model D shows that the model structure of a model will affect the accuracy too and Transformer is better than RNN model.

The accuracy of our model, with 51.9 million parameters, is higher than model C, but the accuracy is close. That means that about 50 million parameters are sufficient. The is to say, the feature vector $v$, described in 2.2, are sufficient, which can be used to current tasks. It needs to be test that if the model can be used in more tasks.

# 6. Conclusion

We designed multi-task pre-trained DNN model for seismic single station waveform processing. The model can be used to Pg/Sg/Pn/Sn phase picking and P polarization classification. The model can be also used to seismic event classification by transfer learning from a small number of samples. The main conclusion are as follows:

1. Our research shows that Pn can be picked from continuous waveform. With longer length of input waveform to train, we can significantly lower the false positive rate, when picking continuous waveform, compared with PhaseNet. The model can be directly used to pick Pg, Sg and Pn phases on continuous waveform, whose accuracy are over 65%. However, the detection accuracy of Sn is low.

2. The accuracy of P polarity classification is over 80%. Our model allows Pg phases located in any position of the 102.4 seconds time window. However, the model designed by Li et al., (2023) will suffer accuracy decrease, when there is time shift of a P phase.

3. Transfer learning is an important goal of our pre-trained model. We used seismic event classification task to test the performance of our pre-trained model. We only trained the type decoder parameters. There are only 7k parameters in type decoder, which means that we need few samples to train decoder.

Our model can be used to many other tasks by means of transfer learning. However, the accuracy of Sn and P polarity still need to be promoted.

**Acknowledgement**

The research is financial supported by National Natural Science Foundation of China (NSFC Grant Number: 4220040580). Source code is available at https://github.com/cangyeone/prime.